# Brief encounter networks

Vassilis Kostakos,* Eamonn O'Neill,* Alan Penn+

* Department of Computer Science, University of Bath, Bath BA2 7AY, UK. (email {vk,eamonn}@cs.bath.ac.uk)
+ The Bartlett, University College London, UK, WC1E 6BT (email a.penn@ucl.ac.uk)

**Many complex human and natural phenomena can usefully be represented as networks describing the relationships between individuals[1,2,3,4]. While these relationships are typically intermittent, previous research has used network representations that aggregate the relationships at discrete intervals[5]. However, such an aggregation discards important temporal information, thus inhibiting our understanding of the network's dynamic behaviour and evolution. We have recorded patterns of human urban encounter using Bluetooth technology (Figure 1) thus retaining the temporal properties of this network. Here we show how this temporal information influences the structural properties of the network. We show that the temporal properties of human urban encounter are scale-free, leading to an overwhelming proportion of brief encounters between individuals. While previous research has shown preferential attachment to result in scale-free connectivity in aggregated network data[11], we found that scale-free connectivity results from the temporal properties of the network. In addition, we show that brief encounters act as weak social ties[6,7] in the diffusion of non-expiring information, yet persistent encounters provide the means for sustaining time-expiring information through a network.**

Our earlier work indicates that about 7.5% of observed pedestrians carry discoverable Bluetooth devices in the city of Bath[8], giving us an approximation of the proportion of the public our technique captures. In Table 1 we list the structural properties of our observation data (Bath), as well as four distinct subsets. The networks exhibit small average paths $\lambda$ and high clustering coefficients $C$, indicative of small-world networks[9]. Furthermore, the Pareto distribution[10] of degree $P(k)$ across the whole dataset follows an approximate power law with $\alpha-1 \approx 1.5$ (Figure 2a), which is characteristic of scale-free networks[11]. Finally, clustering in our dataset follows the approximate relationship $C(k) \approx 1/k$ (Figure 2b), which suggests an underlying modularisation of our data[12,13]. The structural properties of the sub-networks tell a story which intuitively makes sense. For instance we observe the highest $C$ and smallest $\lambda$ in the office rather than the street. Correspondingly, we would expect an office environment to be much more clustered than, say, the street. Furthermore, we see that the street network has the lowest density, while the campus network, being of much smaller size than the street network, has double the number of edges.

Although an examination of the structural properties of encounter networks can provide interesting insights, an aggregated network representation discards valuable temporal information. Techniques have been developed to describe the dynamics of complex networks such as the Brazilian soccer network[2], online dating networks[3] and student affiliation networks[4]. However, such work typically relies on the analysis of a limited number of discrete snapshots of the complex networks.[5] Our data, on the other hand, consists of a chain of events that allows for a minute-by-minute evolving description of the network as people move into and out of contact with each other and our scanners (Figure 1). Here we explore the temporal properties of our network by focusing on three key aspects: presence and frequency of nodes, presence and frequency of links, and temporal order of events.

While in Figure 1c all nodes are visible, in fact they were available sporadically, and only when the corresponding individuals were at a scanning location. We call this availability node presence ($n_p$), calculated as the total amount of time an individual spent near one of our scanners during the study. In Figure 3a (black solid line) we see that $n_p$ follows a power law with $\alpha-1 \approx 0.9$. Thus, whilst most individuals were seen only for a few seconds, others accumulated a presence of more than a month during their visits near our scanners. A further temporal aspect, node frequency ($n_f$), describes the number of distinct instances a person came near a scanner. In Figure 3b we see that $n_f$ follows a power law with $\alpha-1 \approx 1.6$. Thus, most individuals were seen only once, while others were seen on more than a thousand occasions. Finally, we observed that $n_p$ and $n_f$ are not correlated.

When considering human encounters, node availability is a prerequisite for establishing links: a person gains links by being near a scanner and "waiting" for others to show up. Additionally, this attachment is not driven by degree $k$: an individual is attached to *all* other co-present individuals, regardless of their $k$ (as described in Figure 1b). People's behaviour, however, varies in $n_p$ and $n_f$ – some are more "persistent" than others, and thus are more likely to gain new links. For example, a customer entering the pub is more likely to encounter the barman, not because the barman has had many encounters, but because the barman is more likely to actually be in the pub. This is a form of attachment driven by nodes' temporal availability, rather than connectedness per se.

We observe the effect of $n_p$ and $n_f$ by exploring the temporal properties of encounter: link presence ($l_p$) and link frequency ($l_f$). In Figure 3c we see $l_p$ follows a power law with $\alpha-1 \approx 1.3$. Thus, whilst during our study most individuals had only brief encounters, some accumulated a total of up to a few days next to each other during their various encounters. Similarly, in Figure 3d we see $l_f$ follows a power law with $\alpha-1 \approx 1.7$. Thus, whilst most individuals met only once or twice, some met up to 300 times.

To test our hypothesis that node presence and frequency drive the temporal and structural properties of networks, we developed a growth model driven by $n_p$ and $n_f$. At each step of our simulation, a node is activated with probability $n_f$, and if activated, stays active for a duration of $n_p$ times the number of steps it was previously inactive. In our model, attachment between nodes is non-preferential: all simultaneously active nodes are linked to each other. We ran our simulations with varying population sizes, each time resulting in a scale free distribution of $k$, $l_p$ and $l_f$. In Figure 3 (blue dashed line) we see the resulting scale-free distributions of presence and frequency of a simulated population of 70000 nodes using the $n_p$ and $n_f$ derived from our observations, while in Figure 4 we see the resulting scale-free distribution of $k$. Our results suggest that unlike certain networks where nodes are constantly available (e.g. the electricity grid or the road network), when we consider human encounters availability follows a scale-free distribution, which determines the distribution of encounters, or links, between individuals.

In addition to presence and frequency, we explored the temporal order of encounters. Using our data we created an emulation environment in which we studied the diffusion patterns of information packets or digital viruses. In our system, each distinct encounter provides an opportunity for one device to transmit information, or a virus, to another device. Furthermore, to the extent that *people* carry Bluetooth devices, we might begin to extrapolate our findings to physical encounters and biological viruses, although the accuracy of such an extrapolation is uncertain. We emphasise that our system is an emulation, as opposed to a simulation environment: the underlying mechanisms are not probabilistic, but reflect real-world events as recorded by our Bluetooth scanners. Our emulation environment offers a dynamic bond percolation model[14], since we can model the precise time and duration of each encounter.

We carried out exhaustive emulations by injecting the devices with information at all possible points in the chain of events described by our data. We ran these emulations using our complete dataset, as well as subsets where we selectively removed the most brief, or most persistent encounters. In Figure 5 we show the number of devices that non-expiring information can reach over time. We observed that the removal of brief encounters significantly diminishes the ability of non-expiring information to spread through the network, while the removal of persistent encounters has a much smaller effect. This suggests that brief encounters act as weak social ties[6], as they are central to network cohesion[7].

We next injected devices with short-expiry information packets, and observed their diffusion by varying the amount of time they remain active in a device. Such packets may represent time-sensitive messages or viruses, as they remain active in their hosts for a few days. Again we selectively removed the most brief or most persistent encounters from our emulations, and observed the propagation of the packets. In Figure 6 we show the results of our emulations using the Susceptible-Infected-Susceptible model[15] with information packets that remain active for 3 days. In contrast to non-expiring information, we found that the diffusion of time-expiring information lasts longer on networks with persistent encounters.

In this paper we examine the temporal nature of human encounter networks, whose structural properties we found to be both scale-free and small world. Here we show that the temporal properties of human encounter are also scale-free, and we characterise such networks as brief encounter networks due to the overwhelming proportion of short and infrequent encounters between individuals. In brief encounter networks, we show that the temporal behaviour of nodes gives rise to scale-free connectivity with no need of incorporating preferential attachment mechanisms. Furthermore, we demonstrate that brief encounters are fundamental to the propagation of non-expiring information, suggesting that they act as weak social ties between individuals.[6] However, we found that persistent encounters are needed to sustain time-expiring information through a network. Thus, while brief encounters help the spread of gossip and innovations, persistent encounters are crucial in disseminating information such as coordination feedback, traffic information and stock prices.

## Acknowledgements

We thank George Roussos, Gabor Blasko, Matthew Dorrey, Albert-László Barabási and Neil Ferguson for their help and insightful comments.
## References

1. Strogatz, S.H. Exploring complex networks. *Nature*, **410**, 268-276, 2001.

2. Onody, R.N. and de Castro, P.A. Complex network study of Brazilian soccer players. *Physical Review E*, 037103 (2004).

3. Holme, P. Network dynamics of ongoing social relationships. *Europhysics Letters*, **64**, 3, 427-433 (2003).

4. Holme, P., Park, S.M., Kim, B.J., Edling, C.R. Korean university life in a network perspective: Dynamics of a large affiliation network. *Physica A*, **307**, 821-830 (2007).

5. Snijders, T.A.B. The statistical evaluation of social network dynamics. In *Sociological Methodology* (M.E. Sobel and M.P. Becker Eds), Boston and London: Basil Blackwell, 361-395 (2001).

6. Granovetter, M. The strength of weak ties. *American Journal of Sociology*, **78**, 1360-1380 (1973).

7. Bohannon, J. Tracking people's electronic footprints. *Science,* **314**, 914-916 (2006).

8. O'Neill, E., Kostakos, V., Kindberg, T., Fatah gen. Schiek, A., Penn, A., Stanton Fraser, D. and Jones, T. Instrumenting the city: developing methods for observing and understanding the digital cityscape. In proceedings of UbiComp 2006, LNCS 4206, Springer, 315-332 (2006).

9. Watts, D.J. and Strogatz, S.H., Collective dynamics of small-world networks. *Nature*, **393**, 440 (1998).

10. Newman, M.E.J. Power laws, Pareto distributions and Zipf's law. *Contemporary Physics,* **46**, 323-351 (2005).

11. Barabási, A.-L., Albert, R. Emergence of scaling in random networks. *Science* **286**, 509-512 (1999).

12. Dorogovtsev, S.N., Goltsev, A.V. and Mendes, J.F.F., Pseudofractal scale-free web. *Physical Review E*, **65**, 066122 (2002).

13. Ravasz, E., Somera, A.L., Mongru, D.A., Oltvai, Z.N., Barabási, A.-L. Hierarchical organization of modularity in metabolic networks. *Science* **297**, 1551-1555 (2002).

14. Newman, M.E.J., Jensen, I. and Ziff., R.M. Percolation and epidemics in a two-dimensional small world. *Physical Review E*, **65**, 021904 (2002).

15. Anderson, R.M. and May, R.M., Infectious diseases in humans. Oxford University Press, Oxford (1992).

**Figures & Tables**

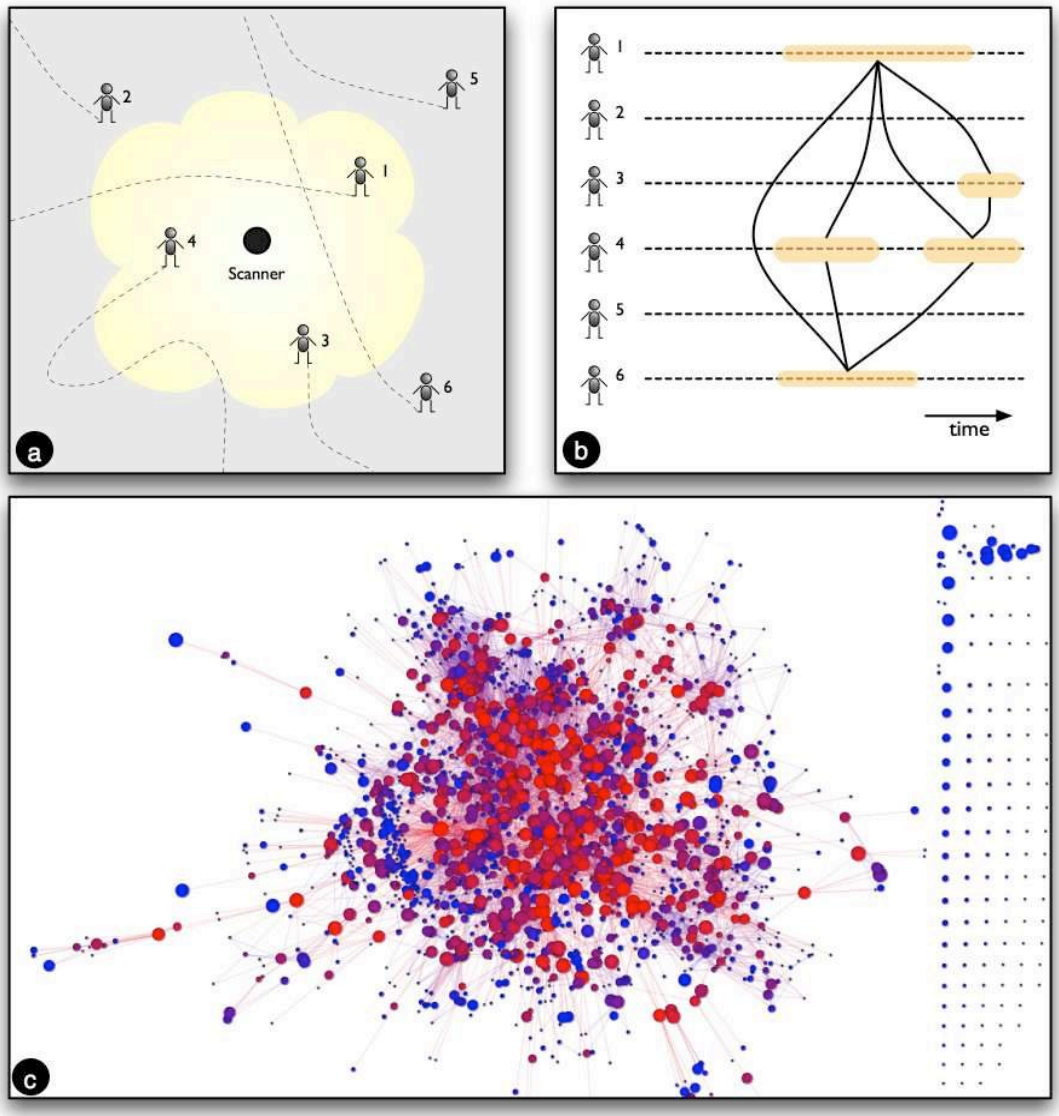

**Figure 1** Deriving networks of copresence from Bluetooth scans over a one-year period. **a**, At each scanning location, our computer uses Bluetooth to monitor the presence of mobile devices within an approximate 10 meter radius. **b**, Each recorded device is allocated its own timeline (dotted horizontal lines). Using data from a, we can plot each device's visit sessions (yellow bars). Overlapping sessions are identified and linked (solid lines). **c**, A network graph is derived by drawing a unique vertex for each device identified in b, and linking those devices that have been linked in b. Because each device transmits a unique ID, we can trace the same device across multiple locations. The graph here depicts the pub network (see Table 1) after 6 months of scanning. Nodes have been sized according to $n_p$, while both vertices and edges are coloured for betweenness (blue to red).

|  | Size | Edges | Density | Core | $k$ | $\lambda_{max}$ | $\lambda$ | $C$ |
|---|---|---|---|---|---|---|---|---|
| Bath | 70516 | 652446 | 0.03% | 69655 | 18.53 | 11 | 3.45 | 0.47 |
| Campus | 3109 | 120273 | 2.5% | 3101 | 77.37 | 6 | 2.57 | 0.44 |
| Street | 11853 | 58111 | 0.08% | 10584 | 9.80 | 12 | 3.23 | 0.28 |
| Pub | 13476 | 126768 | 0.1% | 13383 | 18.81 | 9 | 2.61 | 0.10 |
| Office | 321 | 2419 | 4.7% | 318 | 15.21 | 4 | 2.04 | 0.82 |

**Table 1** Structural properties of the obtained network and some of its subsets. For each subset we show size of the graph, number of edges in the graph, density of edges, size of largest component (core), average degree (*k*), diameter of largest component ($\lambda_{max}$), average length ($\lambda$), and average clustering coefficient (*C*).

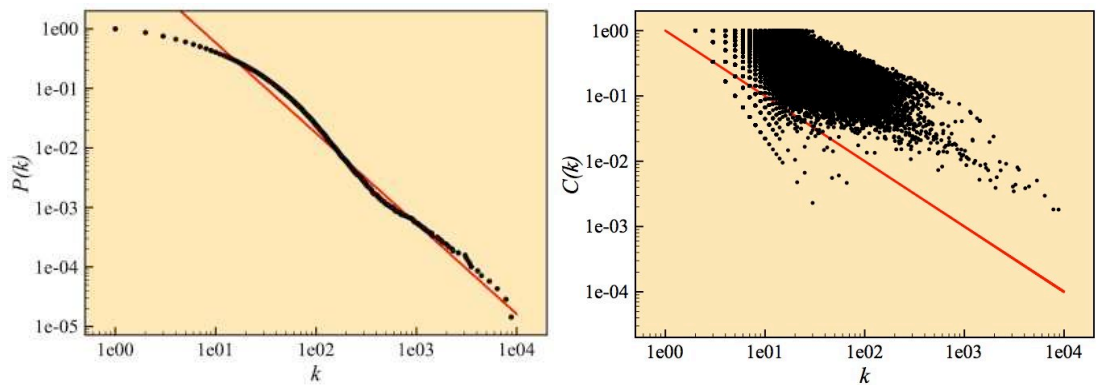

**Figure 2** Structural properties. On the left (Figure 2a), the Pareto distribution of degree *k* follows an approximate power law with $\alpha-1 \approx 1.5$. On the right (Figure 2b), we observe that degree *k* versus clustering coefficient *C(k)* follows the relationship $C(k) \approx 1/k$.

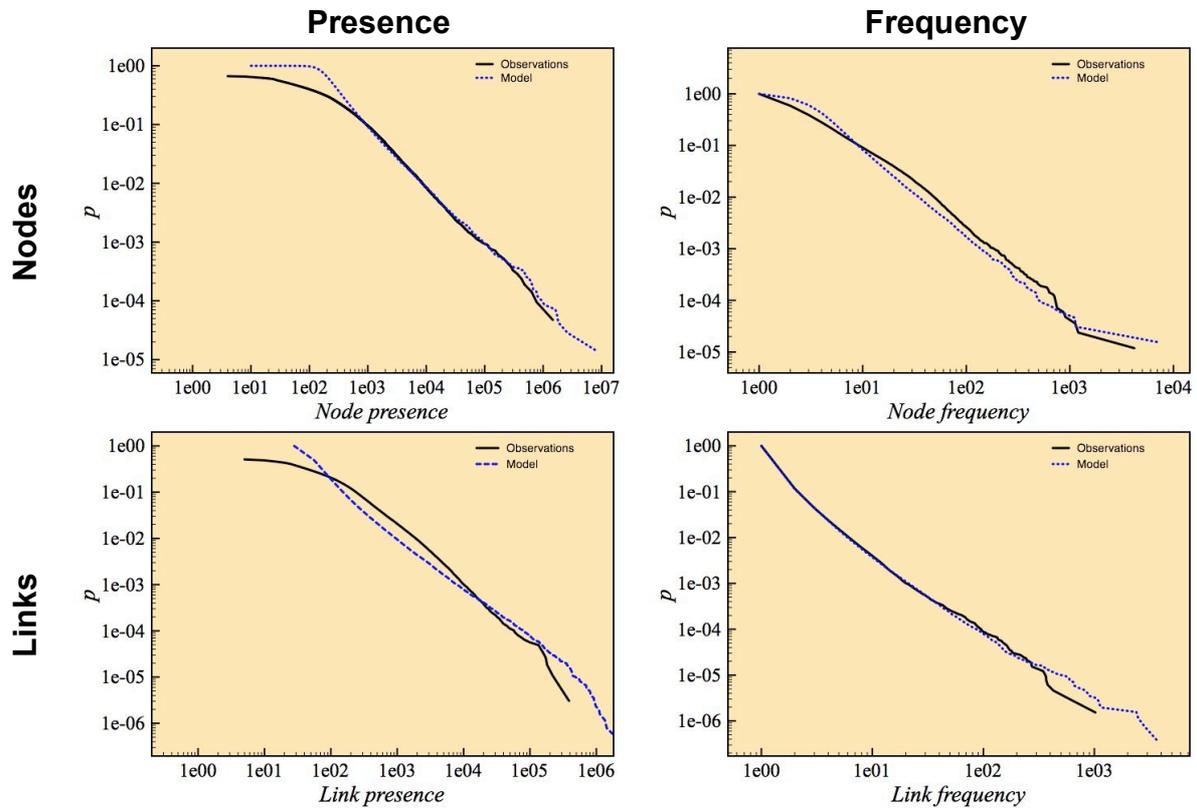

**Figure 3** Temporal properties. Figure 3a (top left) shows the distribution node presence follows a power law with $\alpha-1 \approx 0.9$. Figure 3b (top right) shows the distribution of node frequency follows a power law with $\alpha-1 \approx 1.6$. Figure 3c (bottom left) shows link presence follows a power law with $\alpha-1 \approx 1.3$. In Figure 3d (bottom right), we observe that link frequency follows power law with $\alpha-1 \approx 1.7$.

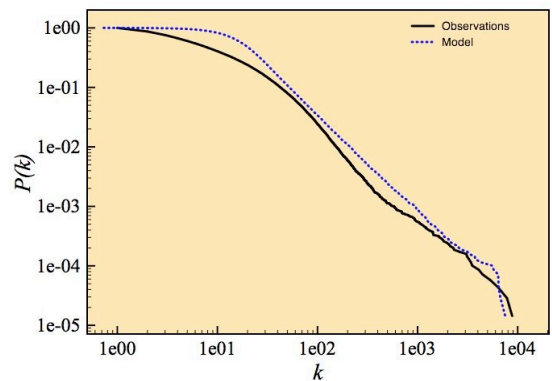

**Figure 4** Degree distribution. Our model (blue dashed line) results in a degree distribution that approximates our observations (black solid line).

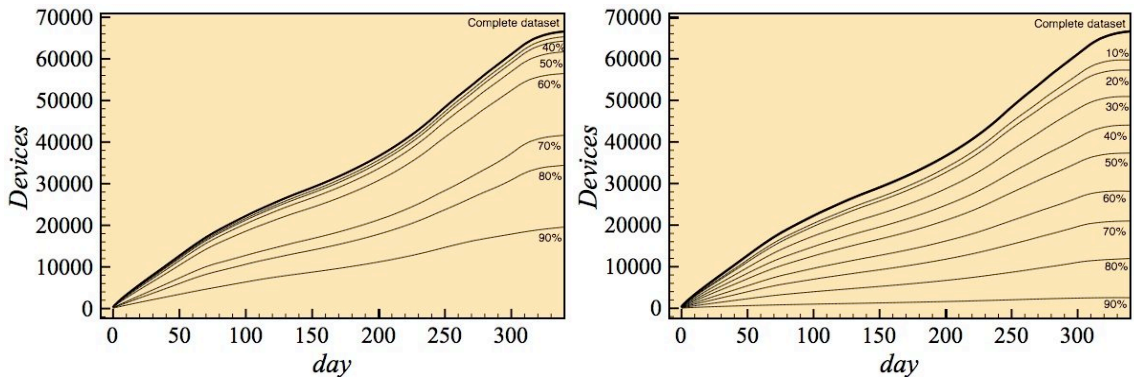

**Figure 5** Information diffusion. Using our emulation environment, we observe the number of devices that information can reach over time using a transmission rate of 1. Additionally, we observe the effect of selectively removing encounters from our dataset. Figure 5a (left) shows the effect of selectively removing the most persistent encounters as a percentage of total encounters, while in Figure 5b (right) we preferentially remove the briefest encounters.

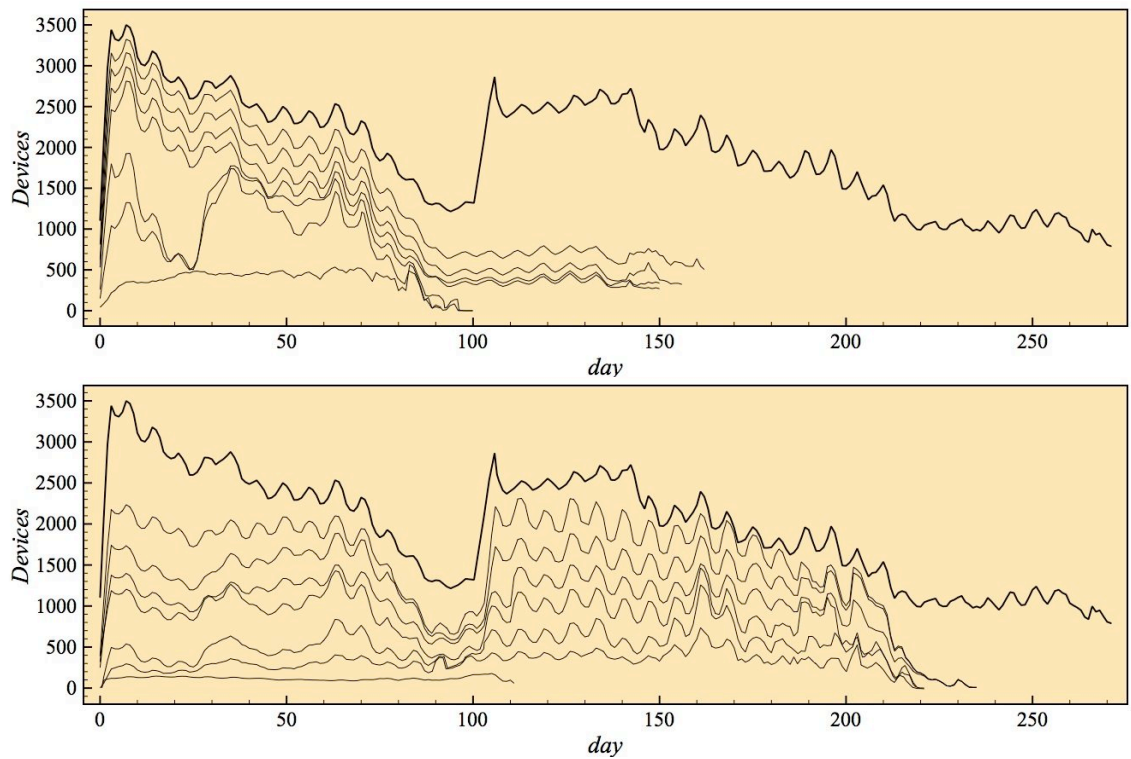

**Figure 6** Virus propagation. A 3-day virus (transmission rate of 1) is injected in our dataset (Figure 6a top), and we selectively remove the most persistent encounters (in steps of 10% of total encounters). In Figure 6b (bottom) we follow the same procedure, selectively removing brief encounters.